# Magnetic and neutron spectroscopic properties of the tetrameric nickel compound [Mo$_{12}$O$_{28}$(μ$_2$-OH)$_9$(μ$_3$-OH)$_3${Ni(H$_2$O)$_3$}$_4$]·13H$_2$O

A. Furrer[1], K.W. Krämer[2], Th. Strässle[1], D. Biner[2], J. Hauser[2], and H.-U. Güdel[2]

[1]Laboratory for Neutron Scattering, ETH Zurich and Paul Scherrer Institut, CH-5232 Villigen PSI, Switzerland

[2]Department of Chemistry and Biochemistry, University of Bern, CH-3012 Bern, Switzerland

**Abstract:**

We present results of inelastic neutron scattering experiments performed for the compound [Mo$_{12}$O$_{28}$(μ$_2$-OH)$_9$(μ$_3$-OH)$_3${Ni(H$_2$O)$_3$}$_4$]·13H$_2$O, which is a molecular magnet with antiferromagnetically coupled Ni$^{2+}$ ions forming nearly ideal tetrahedra in a diamagnetic molybdate matrix. The neutron spectroscopic data are analyzed together with high-field magnetization data (taken from the literature) which exhibit four steps at non-equidistant field intervals. The experimental data can be excellently described by antiferromagnetic Heisenberg-type exchange interactions as well as an axial single-ion anisotropy within a distorted tetrahedron of Ni$^{2+}$ ions characterized by X-ray single-crystal diffraction. Our analysis contrasts to recently proposed models which are based on the existence of extremely large biquadratic (and three-ion) exchange interactions and/or on a strong field dependence of the Heisenberg coupling parameters.

# I. INTRODUCTION

Magnetic molecules consisting of a small number of exchange-coupled paramagnetic ions, which form magnetic clusters embedded in a matrix, have been intensively investigated in recent years. The interest in these systems is twofold: they are ideal model systems to explore the fundamental principles of nanomagnets, and they have a promising application potential as the smallest nanomagnetic units capable of storing information on the molecular level [1]. While the electronic states of the majority of magnetic cluster systems are well understood in terms of common models [2], the consistent interpretation of the magnetic properties of $Ni^{2+}$ tetramers embedded in the molecular compound $[Mo_{12}O_{28}(\mu_2\text{-OH})_9(\mu_3\text{-OH})_3\{Ni(H_2O)_3\}_4]\cdot 13H_2O$ (henceforth abbreviated as $\{Ni_4Mo_{12}\}$) has been controversial [3-5].

Schnack et al. [3] characterized the compound $\{Ni_4Mo_{12}\}$ by bulk magnetic, electron paramagnetic resonance, and magneto-optical experiments. The data were interpreted on the basis of a model involving bilinear and biquadratic exchange interactions as well as a single-ion anisotropy, with all terms being of a similar size; in addition, they proposed a strong field dependence of the bilinear exchange parameters, which are roughly doubled when increasing the magnetic field up to 50 T. Kostyuchenko [4] came up with an alternative model involving bilinear, biquadratic and three-spin exchange interactions, with the latter two interactions being of the order of 12% of the dominant (bilinear) Heisenberg term. Recently, Nehrkorn et al. [5] performed inelastic neutron scattering (INS) experiments and tried to analyze the observed energy spectra by the models used in Refs 3 and 4, however, none of the proposed parameter sets was able to satisfactorily reproduce the data. Hence it was concluded that the conventional spin Hamiltonian approach is not adequate to describe the magnetism in $\{Ni_4Mo_{12}\}$.

We think that the models proposed in Refs 3 and 4 as well as the conclusion of Ref. 5 require revision. Biquadratic exchange indeed exists as an intrinsic second-order mechanism, but the ratio of biquadratic to bilinear exchange is estimated in Anderson's theory of superexchange [6] to be of the order of 1% as widely confirmed by experiments [7]. The sizes of the biquadratic (and three-spin) exchange interactions reported in Refs 3 and 4 exceed this limit by more than an order of magnitude. Furthermore, the strong field dependence of the exchange parameters proposed in Ref. 3 is rather unrealistic. We will show here that the magnetic properties of the $Ni^{2+}$ tetramers in the compound $\{Ni_4Mo_{12}\}$ can be described by the conventional Heisenberg model and an axial single-ion anisotropy. Our considerations are based on the bulk magnetic data reported in Ref. 3 as well as on new results of X-ray diffraction and inelastic neutron scattering experiments which were performed in the course of the present work. The most important point is to realize the subtle but relevant intrinsic deviations of the $Ni^{2+}$ tetramers from an ideal tetrahedral geometry, so that the bilinear exchange interactions between the $Ni^{2+}$ ions can no longer be described by a single exchange parameter J.

The present work is organized as follows. Section II provides details of the sample preparation, the experimental procedures as well as the experimental data. The theoretical background is presented in Sec. III, followed by the data analysis in Sec. IV. Finally, a brief discussion and some conclusions are given in Sec. V.

## II. EXPERIMENTAL DATA

### A. *Chemical structure and sample synthesis*

For the present experiments, deuterated samples of $\{Ni_4Mo_{12}\}$ were synthesized according to Ref. [8]. $(NH_4)_6[Mo_7O_{24}] \cdot 4H_2O$ (6.72 g, 5.437 mmol, Fluka p.a.) and $Ni(CH_3COO)_2 \cdot 4H_2O$ (22.50 g, 90.412 mmol, Fluka p.a.) were dissolved in 400 ml $D_2O$ (ARMAR, 99.8% D) and 40 ml acetic acid (Fluka p.a.) in a flask under argon. Hydrazinium sulfate (1.24 g, 57.178 mmol, Fluka p.a.) was added at room temperature and stirred until dissolution. Then the solution was warmed to 50°C for 1 day and 65°C for 2 days. Brown crystals were filtered off and kept under $D_2O$. From the D/H ratio of the reaction mixture the sample was 96.4% deuterated.

The product contained brown plate-like crystals of $\{Ni_4Mo_{12}\}$ and additionally about 10% dark brown crystals of pseudo-octahedral shape. The latter ones were identified by single crystal X-ray diffraction as $H_{10}[NiMo_{12}O_{40}\{Ni(H_2O)_3\}_4]$ [9], in the following abbreviated as $\{Ni_5Mo_{12}\}$. Since $\{Ni_5Mo_{12}\}$ formed bigger crystal aggregates, part of this byproduct could be removed by sieving.

$\{Ni_4Mo_{12}\}$ is very sensitive to dehydration. Already at room temperature water is readily lost and crystalline samples convert into a nano-crystalline powder. The dehydration can be followed by powder X-ray diffraction. On drying, the narrow diffraction peaks of a crystalline sample turn into very broad peaks of a close-to-amorphous product. A thermo-gravimetric measurement revealed the main water loss between room temperature and 100°C, followed by a continuous slower loss of weight up to 400°C. A scanning electron microscope image shows the shrunk, cracked crystals due to loss of crystal water in the vacuum of the instrument.

For the X-ray structure determination a crystal plate of size 0.067 x 0.061 x 0.020 mm$^3$ was mounted in a MiTeGen loop using Paratone-N on a Bruker APEXII CCD Diffractometer System (Bruker, 2007) equipped with a three-circle goniometer and a graphite-monochromator. The data collection was performed at T=183 K using Mo-K$_\alpha$ radiation ($\lambda$=0.71073 Å). 1224 exposures (60 seconds per exposure) were obtained at a detector distance of 49.5 mm with one $\omega$-scan and two $\varphi$-scans and the crystal oscillating through 0.5°. The structure was solved by direct methods using the program SHELXS-97 [10] and refined by a full matrix fitting procedure applied to the squares of the structure factor with SHELXL-97 [11]. The resulting structure parameters are listed in Ref. 12. The crystal water molecules are partially disordered and not well localized, which results in rather big thermal displacement parameters. The hydrogen atoms could not be located in the difference electron density map. The highest remaining electron density is 1.77 electrons/Å$^3$.

The [Mo$_{12}$O$_{28}$($\mu_2$-OH)$_9$($\mu_3$-OH)$_3${Ni(H$_2$O)$_3$}$_4$] molecule is shown in Fig. 1. It contains 4 Ni$^{2+}$ ions located at the corners of a distorted tetrahedron. The cluster has two shorter, i.e. Ni2-Ni3 and Ni2-Ni3', and four longer Ni-Ni distances, see Table I. This is in contrast to Ref. 8 where a structure with three shorter and three longer Ni-Ni distances is reported. The Ni$^{2+}$ ions are coordinated by three H$_2$O molecules and three O$^{2-}$ ions of [Mo$_2$O$_{10}$] groups, see Fig. 2a. All Mo atoms in {Ni$_4$Mo$_{12}$} are Mo$^V$. Accordingly, the [Mo$_2$O$_{10}$] groups contain a Mo$^V$-Mo$^V$ single bond and are diamagnetic. Each Mo is surrounded by six oxygen atoms in the shape of a distorted octahedron. The dimolybdate groups are located above the edges of the tetrahedron, see Fig. 2b, and further condensed to a $[Mo_{12}O_{40}H_{12}]^{8-}$ cluster. Its charge is balanced by four $Ni^{2+}$ and twelve $H^+$ ions. The $H^+$ ions could not be localized by X-ray diffraction. Their positions can be inferred, however, as will be discussed in Sec. IV.A. They reduce the cluster symmetry from $T_d$ to $C_s$ and are thus crucial for understanding the exchange interaction.

Aside from the terminal water ligands of the $Ni^{2+}$ ions, the oxygen atoms of the cluster are divided into four groups. Each Mo has one terminal $\mu_1$-O atom, located above the drawing plane in Fig. 2b. $\mu_2$-O atoms bridge between $[Mo_2O_{10}]$ groups. They are located in the drawing plane of Fig. 2b, and there are twelve of them. Both $\mu_1$-O and $\mu_2$-O atoms are located on the surface of the cluster. $\mu_3$-O atoms connect between three $[Mo_2O_{10}]$ groups. They are located below the drawing plane in Fig. 2b, and there are four of them. A second type of $\mu_3$-O atoms connects a Ni and two Mo atoms, and they add up to twelve within the cluster. Relevant angles and distances are listed in Table I.

*B. Bulk magnetic data*

Magnetic susceptibility and high-field magnetization data up to 60 T were performed by Schnack et al. [3] on a powder sample of $\{Ni_4Mo_{12}\}$. Both measurements show the characteristics of a singlet ground-state system, so that the coupling within the $Ni^{2+}$ tetramers is antiferromagnetic. Of particular interest are the magnetization data taken at T=0.44 K, which exhibit four steps before reaching saturation; this observation points to the occurrence of ground-state level crossings [3]. The four steps were found at 4.5, 8.9, 20.1, and 32 T, according to the maxima of the differential magnetization data displayed in Fig. 3. With increasing field, the magnetization steps are increasingly smoothed out but can still be located. Most importantly, the four steps occur at non-equidistant field intervals.

## C. Inelastic neutron scattering (INS) experiments

The INS experiments were carried out with use of the high-resolution time-of-flight spectrometer FOCUS at the spallation neutron source SINQ at PSI Villigen. The measurements were performed with incoming neutron energies of $E_0$=5.93 and 3.27 meV, giving rise to energy resolutions at the elastic position of 0.3 and 0.1 meV, respectively. The scattered neutrons were detected by an array of $^3$He counters covering a large range of scattering angles 10°≤Φ≤130° corresponding to moduli of the scattering vector up to Q≈2.5 Å$^{-1}$ (for $E_0$=5.93 meV). A fresh crystalline sample was removed from the mother liquor, washed with a small amount of $D_2O$, and then immediately filled into an Al cylinder (10 mm diameter, 45 mm height) and placed into a He cryostat to achieve temperatures in the range 1.5 < T < 30 K. Part of the impurity phase {Ni$_5$Mo$_{12}$} was removed from the batch by sieving before sealing the sample in the container. The raw data were corrected for background, detector efficiency, absorption, and detailed balance effects according to standard procedures.

Energy spectra of neutrons ($E_0$=5.93 meV) scattered from polycrystalline {Ni$_4$Mo$_{12}$} are shown in Fig. 4. The 4% hydrogen content of the sample gives rise to a strong incoherent elastic peak, which prevents the observation of resolved lines below energy transfers ΔE≈1 meV. The T and Q dependences of the intensities for ΔE>2 meV are to a large extent characteristic of phonon scattering. For ΔE<2 meV, on the other hand, the T and Q dependence is typical of magnetic excitations. This is confirmed by the energy spectra taken with improved energy resolution ($E_0$=3.27 meV) as shown in Fig. 5a, which unravels further details for $\Delta E$<1 meV. In Fig. 5b we reproduce the high-resolution INS data from Ref. 5. The comparison of our data with those of Ref. 5 clearly shows that they are very similar with one exception. The peak centered at 1.35 meV in our spectra is absent in Fig. 5b. We therefore suspect that it is due to an impurity in our sample. This is confirmed by examining the T and Q dependence of our INS data, see Figs 6 and 7. The T dependence of the intensities is displayed in Fig. 6 and compared to the Boltzmann population factors of the ground state ($n_0$) and the first-excited state ($n_1$) associated with the energy-level sequence of the $Ni^{2+}$ tetramers (see Section IV.C). The agreement between the observed and calculated intensities is excellent except for the peak at 1.35 meV. Fig. 7 shows the Q dependence of our INS data. The intensities outside the range 1.2-1.5 meV follow the oscillatory Q dependence (Fig. 7a) predicted by the magnetic neutron cross section for a polycrystalline sample containing exchange-coupled tetrahedra [13]:

$$I(Q) \propto F^2(Q)[1-\sin(QR)/QR] \qquad (1)$$

Here, F(Q) is the magnetic form factor, and R denotes the intracluster Ni-Ni distance. The peak centered at 1.35 meV, on the other hand, does not follow this oscillation (Fig. 7b). We conclude that the feature at 1.35 meV is indeed an impurity band, and in the quantitative analysis the data points between 1.2 meV and 1.5 meV will be left out. With this correction, our data are in excellent agreement with those of Ref. 5 (Fig. 5b), which were obtained on a different spectrometer and likely with a partially dehydrated sample.

Ignoring the impurity, we can identify two overlapping cold bands at about 0.42 meV and 0.60 meV. Another cold band, which is rather broad, is centered at 1.7 meV. Two warm bands, both rather broad, are present at 1.1 and 2.4 meV. We thus have excited cluster states at 0.42 meV, 0.60 meV and 1.7 meV, which are accessible by INS from the ground state, in very good agreement with the qualitative analysis in Ref. 5.

## III. THEORETICAL BACKGROUND

The spin Hamiltonian of an ideal $Ni^{2+}$ tetramer forming an equilateral tetrahedron is given in the Heisenberg approximation by

$$H = -2J(\mathbf{s}_1\cdot\mathbf{s}_2+\mathbf{s}_1\cdot\mathbf{s}_3+\mathbf{s}_1\cdot\mathbf{s}_4+\mathbf{s}_2\cdot\mathbf{s}_3+\mathbf{s}_2\cdot\mathbf{s}_4+\mathbf{s}_3\cdot\mathbf{s}_4) \quad , \tag{2}$$

where J is an adjustable exchange parameter and $\mathbf{s}_i$ the spin operator of an individual $Ni^{2+}$ ion with $s_i=1$. In the following we modify Eq. (2) by introducing distortions of the tetrahedron, which can be described by a two-parameter model (J, J'). We consider three models in which one, two, and three of the six exchange parameters are different from J as visualized in the inserts of Figures 8 and 9. For model 1 the spin Hamiltonian is described by

$$H_1 = -2J(\mathbf{s}_1\cdot\mathbf{s}_2+\mathbf{s}_1\cdot\mathbf{s}_3+\mathbf{s}_1\cdot\mathbf{s}_4+ \mathbf{s}_2\cdot\mathbf{s}_3+\mathbf{s}_2\cdot\mathbf{s}_4) -2J'\mathbf{s}_3\cdot\mathbf{s}_4 \quad . \tag{3}$$

For a complete characterization of the tetramer states, we need additional spin quantum numbers resulting from the vector sums $\mathbf{S_{12}}=\mathbf{s_1}+\mathbf{s_2}$, $\mathbf{S_{34}}=\mathbf{s_3}+\mathbf{s_4}$, and $\mathbf{S}=\mathbf{S_{12}}+\mathbf{S_{34}}$ with $0 \leq S_{12} \leq 2s_i$, $0 \leq S_{34} \leq 2s_i$, and $|S_{12}-S_{34}| \leq S \leq (S_{12}+S_{34})$, respectively. The eigenvalues of Eq. (3) are independent of the quantum number $S_{12}$:

$$E_1(S_{34},S) = -J\left[S(S+1) - S_{34}(S_{34}+1) - 2s_i(s_i+1)\right] \\ - J'\left[S_{34}(S_{34}+1) - 2s_i(s_i+1)\right] \quad . \tag{4}$$

For model 2a the spin Hamiltonian reads

$$H_{2a} = -2J(\mathbf{s_1}\cdot\mathbf{s_2}+\mathbf{s_1}\cdot\mathbf{s_3}+\mathbf{s_1}\cdot\mathbf{s_4}+\mathbf{s_3}\cdot\mathbf{s_4}) -2J'(\mathbf{s_2}\cdot\mathbf{s_3}+\mathbf{s_2}\cdot\mathbf{s_4}) \quad , \tag{5}$$

which can be brought to diagonal form by choosing the spin quantum numbers $\mathbf{S_{34}}=\mathbf{s_3}+\mathbf{s_4}$, $\mathbf{S_{234}}=\mathbf{s_2}+\mathbf{S_{34}}$, and $\mathbf{S}=\mathbf{s_1}+\mathbf{S_{234}}$ with $0 \leq S_{34} \leq 2s_i$, $|S_{34}-s_i| \leq S_{234} \leq (S_{34}+s_i)$, and $|S_{234}-s_i| \leq S \leq (S_{234}+s_i)$, respectively. The eigenvalues of Eq. (5) are then given by

$$E_{2a}(S_{34},S_{234},S) = -J\left[S(S+1) - S_{234}(S_{234}+1) + S_{34}(S_{34}+1) - 3s_i(s_i+1)\right] \\ - J'\left[S_{234}(S_{234}+1) - S_{34}(S_{34}+1) - s_i(s_i+1)\right] \quad . \tag{6}$$

A variant of model 2a is given by the spin Hamiltonian

$$H_{2b} = -2J(\mathbf{s_1}\cdot\mathbf{s_3}+\mathbf{s_1}\cdot\mathbf{s_4}+\mathbf{s_2}\cdot\mathbf{s_3}+\mathbf{s_2}\cdot\mathbf{s_4}) -2J'(\mathbf{s_1}\cdot\mathbf{s_2}+\mathbf{s_3}\cdot\mathbf{s_4}) \quad , \tag{7}$$

for which the spin quantum numbers used to solve Eq. (3) are appropriate to derive the eigenvalues:

$$E_{2b}(S_{12},S_{34},S) = -J[S(S+1) - S_{12}(S_{12}+1) - S_{34}(S_{34}+1)] \\ - J'[S_{12}(S_{12}+1) + S_{34}(S_{34}+1) - 4s_i(s_i+1)] \quad . \tag{8}$$

Finally, model 3a is described by the spin Hamiltonian

$$H_{3a} = -2J(s_1 \cdot s_2 + s_1 \cdot s_3 + s_1 \cdot s_4) - 2J'(s_2 \cdot s_3 + s_2 \cdot s_4 + s_3 \cdot s_4) \quad . \tag{9}$$

Here, we adopt the spin coupling scheme used for Eq. (5). The eigenvalues turn out to be independent of the spin quantum number $S_{34}$:

$$E_{3a}(S_{234},S) = -J[S(S+1) - S_{234}(S_{234}+1) - s_i(s_i+1)] \\ - J'[S_{234}(S_{234}+1) - 3s_i(s_i+1)] \quad . \tag{10}$$

A variant of model 3a is given by the Hamiltonian

$$H_{3b} = -2J(s_1 \cdot s_3 + s_1 \cdot s_4 + s_2 \cdot s_4) - 2J'(s_1 \cdot s_2 + s_2 \cdot s_3 + s_3 \cdot s_4) \quad , \tag{11}$$

for which none of the above coupling schemes results in a diagonal energy matrix, so that the eigenvalues $E_{3b}$ have to be calculated by conventional spin operator techniques. Figs 8 and 9 display the energy levels E(S) normalized to J (assuming antiferromagnetic coupling J<0) as a function of the ratio x=J'/J. The degeneracies of the spin states |S> are completely lifted only for model 3b as shown in Fig. 9. For x=1 the energy level splittings follow the Landé interval rule E(S)-E(S-1)=-2JS for all models.

The requested singlet ground state (S=0) is realized for all parameters x>0, the only exception being model 3a for which a singlet ground state is restricted to the parameter range 0.5<x<2.

## IV. ANALYSIS OF THE EXPERIMENTAL DATA

### A. *Composition and structure*

In the original paper [8] the title compound was formulated as $[Mo_{12}O_{30}(\mu_2\text{-OH})_{10}H_2\{Ni(H_2O)_3\}_4]\cdot 14H_2O$. The complexes in the square bracket are embedded in a sea of crystal water, which is largely disordered. In our structure determination we find 13 $H_2O$ molecules and a unit cell that is 0.8% smaller in volume than in Ref. 8 . Left on the bench, all or part of the crystal water is lost. We studied freshly prepared crystalline samples. The authors of Refs 3 and 5 give no indication about the water content or the crystallinity of the samples used in their experiments. It is reasonable to assume that the samples were partially dehydrated.

The formulation $[Mo_{12}O_{30}(\mu_2\text{-OH})_{10}H_2\{Ni(H_2O)_3\}_4]$ of the complex already indicates, that the twelve molybdate protons are chemically not equivalent. This is correct, but the above formulation is incompatible with the crystallographic symmetry, which is lowered from $T_d$ to $C_s$. The mirror plane contains the Ni1 and Ni2 positions and bisects the Ni3-Ni3' connection. The following discussion is based on the $C_s$ symmetry and the Mo-O distances and Mo-O-Mo angles determined in the present study (Table I). The $\mu_3$-oxo ions point towards the center of the complex. Their negative charge has to be compensated by placing some positive charge into this central cavity of the cluster. This can be achieved by protons or, alternatively, by the introduction of a cation such as $Ni^{2+}$. In the latter case, the resulting $H_{10}[NiMo_{12}O_{40}\{Ni(H_2O)_3\}_4]$ complex is known and actually occurs as an impurity in our sample, see Sec. II.A. Although we cannot determine H positions from our X-ray data, we confidently conclude that three protons are placed within this cavity, and the only unprotonated $\mu_3$-oxo ion is O14. It is marked in Fig. 1 with a yellow color. The remaining nine protons are situated on the outside of the cluster. Of the twelve $\mu_2$-O bridges connecting the $Mo_2O_{10}$ pairs, nine are $\mu_2$-hydroxo and three $\mu_2$-oxo. Again, based on symmetry and Table I, we assign O19, O20 and O20' as unprotonated $\mu_2$-oxo bridging ions. They are also marked with a yellow color in Fig. 1. The net result is that the Ni2-Ni3 and Ni2-Ni3' bridges are distinctly different from the other four bridges. This is reflected in the shorter Ni2-Ni3 = Ni2-Ni3' distance of 6.663 Å, compared to a range 6.685-6.698 Å for the other four. In Figs 1 and 2a these two Ni-Ni connections are highlighted by black lines. The proper formulation of the complex is thus $[Mo_{12}O_{28}(\mu_2\text{-OH})_9(\mu_3\text{-OH})_3\{Ni(H_2O)_3\}_4]$.

## B. High-field magnetization

The high-field magnetization data observed by Schnack et al. [3] are ideally suited to check the validity of the model spin Hamiltonians outlined in Sec. III. Step-like features occur whenever the Zeeman energy $E_Z=g\mu_B H$ compensates the exchange energy associated with the lowest states of a given S value (S=1,2,3,4) marked with bold lines in Figs 8 and 9, thereby successively increasing the total spin of the Ni tetramer by $\Delta S=1$ from S=0 up to S=4. The exchange energies for models 1, 2a, 2b, and 3a result from Eqs (4), (6), (8), and (10), respectively. The relations between the exchange and Zeeman energies are summarized in Table II [14], with the numbers for model 3b being rounded to one decimal digit. Since the magnetization steps occur at non-equidistant field intervals, we do not consider energy level schemes that are governed by the Landé interval rule (as realized in models 1 and 2b for x<1). The best agreement with the observed data is obtained for model 1 (x>1) with a standard deviation $\chi^2$=0.31, followed by model 2a (x>1) with $\chi^2$=1.77, whereas the other models essentially fail to correctly reproduce the third and fourth magnetization steps. This outcome is actually expected by visual inspection of Fig. 8, which for x>1 shows a Landé splitting pattern for the lowest states S=0, S=1, and S=2, followed by a considerable enhancement of the energy differences to the higher states S=3 and S=4. These requested features are not realized in the energy level patterns displayed in Fig. 9.

## C. Inelastic neutron scattering (INS) experiments

Whereas the high-field magnetization data contain intrinsic information on the complete energy level sequence of the $Ni^{2+}$ tetramers, the energy spectra displayed in Figs 4 and 5 probe primarily the low-energy part. For the analysis of the neutron data we will use the difference of energy spectra taken at different temperatures ($T_1$ and $T_2$), which has the advantage that uncertainties about the background are automatically eliminated. This is exemplified for $T_1$=1.5 K and $T_2$=10 K in Figs 10 and 11a. In principle, positive intensity differences correspond to the presence of ground-state transitions, whereas negative intensity differences result from excited-state transitions. We exclude the range 1.2 < $\Delta E$ < 1.5 meV from our analysis, because the corresponding excitation cannot be associated with the $Ni^{2+}$ tetramers as concluded in Sec. II.C. Similarly, we exclude the low-energy part close to the elastic line, where the difference intensities exhibit an unexpected upturn. The data analysis is based on the neutron cross section defined by Eq. (1), which for each transition i→j has to be individually weighted by the Boltzmann population factor $n_i$ of the initial state as well as the squared transition matrix elements $P_{i \to j}$ [15]:

$$I(Q,T_1) - I(Q,T_2) \propto F^2(Q)[1 - \sin(QR)/QR] \sum_{ij} P_{i \to j} [n_i(T_1) - n_i(T_2)] \quad . \quad (12)$$

In a first approximation, we ignored the splitting of the cold band at 0.5 meV and tested the models presented in Sec. III. Least-squares fitting procedures were applied to a variety of difference energy spectra on the basis of Eq. (12), with the parameters listed in Table II as starting values for all the models. All these attempts failed to produce a quantitative agreement between the observed and calculated difference spectra except for model 2a, which converged to the parameters

$$J = -3.0(2) \text{ K} , \quad J' = -6.2(4) \text{ K} \quad (13)$$

with an intrinsic linewidth of about 0.3 meV. The calculated difference spectra for $T_1$=1.5 K and $T_2$=10 K correctly reproduce both the observed sign changes and the relative strengths of the difference intensities as illustrated in Figs 10 and 11a. The corresponding energy level sequence is displayed in Fig. 12, in which some relevant transitions are marked by arrows. The transitions are governed by the selection rules $\Delta S_{34}$=0,±1;

$\Delta S_{234}=0,\pm1$; $\Delta S=0,\pm1$. There are two dominant ground-state transitions, namely $|2,1,0\rangle \rightarrow |2,1,1\rangle$ and $|2,1,0\rangle \rightarrow |1,1,1\rangle, |2,2,1\rangle$, giving rise to positive intensity differences at energy transfers of 0.5 and 1.7 meV, respectively. The negative intensity differences at energy transfers of about 1 and 2.4 meV are due to the excited-state transitions $|2,1,1\rangle \rightarrow |2,1,2\rangle, |1,1,1\rangle, |2,2,1\rangle$ and $|2,1,1\rangle \rightarrow |1,1,2\rangle, |2,2,2\rangle$, respectively.

The splitting of the band centered at 0.5 meV into two components can be ascribed to anisotropy. We introduced an axial single-ion anisotropy of the form

$$H_D = D\sum_{i=1}^{4}(s_i^z)^2 \quad , \tag{14}$$

which has the effect to split the spin states $|S\rangle$ into the states $|S,\pm M\rangle$ with $-S \leq M \leq S$. Including $H_D$ in the least-squares fitting procedures for model 2a resulted in an improvement of the standard deviation by 10% with the parameters

$$J = -2.9(2) \text{ K}, \quad J' = -6.1(4) \text{ K}, \quad D = 2.6(5) \text{ K} \tag{15}$$

and a slight reduction of the intrinsic linewidth, thereby compensating for the splitting induced by the single-ion anisotropy. The single-ion anisotropy splits the first-excited triplet state $|2,1,1\rangle$ into a doublet and a singlet separated by 0.19 meV, which can also be guessed by visual inspection of the high-resolution energy spectra of Figures 5 and 11.

We extended our least-squares fitting procedures and allowed up to six exchange couplings $J_{ij}$ ($i<j=1,2,3,4$) to vary independently. We realized that in all cases the standard deviation was only marginally improved by 5% at most, and the parameters converged more or less to the restriction set by the model 2a.

As a further test for the validity of our model we also applied it to the data of Ref. 5. The intensity difference of the spectra at $T_1$=2.4 K and $T_2$=9.3 K is displayed in Fig. 11b. A least-squares fit of model 2a including axial anisotropy produced the following parameters:

$$J = -2.9(2) \text{ K} , \quad J' = -6.6(4) \text{ K} , \quad D = 2.3(5) \text{ K} . \qquad (16)$$

They are very close to those determined from our data. We note, in particular, that the intensity of the cold band at 1.7 meV, which caused a lot of problems in Ref. 5, is quantitatively reproduced.

We note that the high-field magnetization data are well reproduced by the model parameters resulting from the INS experiments, see Fig. 3. Our choice of the spectroscopic splitting factor g=2.25 is in agreement with the analysis of the magnetic susceptibility data [3].

At present, the origin of the impurity peak at 1.35 meV is unknown. It may be due to a ground-state excitation of the $Ni^{2+}$ pentamers associated with the minority phase {$Ni_5Mo_{12}$} present in the sample. $Ni^{2+}$ pentamers have a magnetic ground state, giving rise to strong quasielastic scattering, which may also explain the observed upturn of the intensity differences at low energy transfers (see Figs 10 and 11a). The geometry of the $Ni^{2+}$ pentamer is similar to the $Ni^{2+}$ tetramer, with four $Ni^{2+}$ ions placed at the corners of a tetrahedron and the fifth $Ni^{2+}$ ion sitting in the center. The presence of two different intracluster Ni-Ni bond lengths results in a superposition of two different structure factors which largely cancel each other, so that the neutron cross section is governed essentially by the magnetic form factor alone (see Fig. 7b).

## V. DISCUSSION AND CONCLUDING REMARKS

It is remarkable that the analysis of the physical data, in particular the INS data, which was carried out without any assumptions concerning the distortion of the complex, finds model 2a to be the only one compatible with all the data. Model 2a corresponds exactly to the situation derived from the X-ray data and depicted in Fig. 1. The two bridges Ni2-Ni3 and Ni2-Ni3' are different from the other four connections. It is important to realize, that this is an intrinsic property of the $[Mo_{12}O_{28}(\mu_2\text{-OH})_9(\mu_3\text{-OH})_3\{Ni(H_2O)_3\}_4]$ complex and not caused by the packing of the complexes in the crystal. The inequivalence of the Ni-Ni bridges is retained when the crystal is losing water and becomes amorphous. This is very nicely confirmed by the observation that the INS spectra of the crystallized hydrated sample (Fig. 5a) and those taken from Ref. 5 (Fig. 5b) are essentially the same.

In clusters of $Ni^{2+}$ ions, both ferro- and antiferromagnetic couplings have been found, a clear indication that the net J results from an interplay of ferro- and antiferromagnetic orbital contributions. In the title complex, both J and J' are antiferromagnetic, which is reasonable, considering that the superexchange occurs through the dimolybdate network, see Fig. 2. J' is bigger than J by about a factor of two, quite a significant difference. We ascribe this to differences in the electron density distribution along the exchange pathways. Neighboring dimolybdate groups are connected by oxo or hydroxo bridges, as discussed in Sec. IV.A. Electronically, there is a significant difference between an oxo and a hydroxo bridge. In the latter, the proton pulls electron density out of the O-Mo bond, thus reducing the electron density on the respective dimolybdate groups. The distribution of oxo bridges within the complex (yellow atoms in Fig. 1) leads to an overall higher electron density on the dimolybdate groups connecting Ni2-Ni3 and Ni2-Ni3' (J') than along the other (J) exchange pathways. It therefore makes intuitive sense that the kinetic exchange is more efficient and thus J' bigger than J. Our model with two exchange parameters is an approximation. From the structural parameters, $J_{12}$, $J_{13}$ and $J_{33'}$ are not expected to be exactly the same. But since the addition of more parameters did not improve the fit to the INS data, we conclude that model 2a is a good approximation.

In conclusion, the magnetic properties of the compound {$Ni_4Mo_{12}$} can be explained in terms of a conventional model involving Heisenberg exchange interactions and an axial single-ion anisotropy term. This is in contrast to the conclusions of some recent publications, in which the authors proposed rather exotic models or suggested parameter values which are physically not plausible [3-5]. These authors failed to recognize, that the title complex is not and cannot be tetrahedral. This is understandable, because it is not obvious at first sight and requires a careful examination of the chemical structure. The authors of the original work on this compound [8] presented magnetic susceptibility data which could be reasonably reproduced by a tetrahedral model. The magnetisation data up to 60 Tesla [3] and the INS data [5], however, are much more susceptible to the detailed splitting of the energy levels, and more refined models are needed. The authors of the recent INS study [5] obviously felt ill-at-ease with their proposed models. They write "the intensity of peak III will in fact turn out to be a major obstacle for all the models in this work". We now know that there is nothing wrong with peak III, i.e. the intense cold band centered at 1.7 meV, but with the models. In the past 35 years we have studied dozens of exchange coupled clusters of transition metal ions by INS and magnetic measurements. And it is gratifying to find that the title compound [$Mo_{12}O_{28}(\mu_2$-$OH)_9(\mu_3$-$OH)_3\{Ni(H_2O)_3\}_4$]·13$H_2O$ is no exceptional case and can be modeled with a simple Hamiltonian and plausible parameters.

## ACKNOWLEDGMENTS

This work was performed at the Swiss Spallation Neutron Source SINQ, Paul Scherrer Institute (PSI), Villigen, Switzerland. Financial support by the Swiss National Science Foundation is gratefully acknowledged.

**Table I.** Ni-Ni and Mo-O distances and angles of {Ni$_4$Mo$_{12}$} determined by single crystal X-ray diffraction at T=183 K.

|  | distance [Å] |  | distance [Å] |
|---|---|---|---|
| Ni1 – Ni2 | 6.685(2) | Ni1 – Ni3 (2x) | 6.6980(15) |
| Ni2 – Ni3 (2x) | 6.6626(18) | Ni3 – Ni3' | 6.696(2) |
| μ$_1$-O |  |  |  |
| O6 – Mo1 (2x) | 1.686(6) | O21 – Mo5 (2x) | 1.696(5) |
| O10 – Mo2 (2x) | 1.698(5) | O23 – Mo6 | 1.711(8) |
| O13 – Mo3 (2x) | 1.688(6) | O24 – Mo7 | 1.689(8) |
| O17 – Mo4 (2x) | 1.704(5) |  |  |
| μ$_2$-O |  |  |  |
| O4 – Mo1 (2x) | 2.109(6) | O15 – Mo4 (2x) | 2.122(5) |
| O4 – Mo2 (2x) | 2.128(6) | O15 – Mo6 (2x) | 2.109(5) |
| O5 – Mo1 (2x) | 2.102(6) | O16 – Mo4 | 2.107(4) (2x) |
| O5 – Mo3 (2x) | 2.093(6) | O19 – Mo5 | 2.093(4) (2x) |
| O9 – Mo2 (2x) | 2.122(5) | O20 – Mo5 (2x) | 2.065(6) |
| O9 – Mo3 (2x) | 2.093(5) | O20 – Mo7 (2x) | 2.070(5) |
| μ$_3$-O |  |  |  |
| O2 – Mo1 (2x) | 2.187(5) | O14 – Mo4 | 2.149(4) (2x) |
| O2 – Mo2 (2x) | 2.135(5) | O14 – Mo6 | 2.092(7) |
| O2 – Mo3 (2x) | 2.211(5) | O18 – Mo5 | 2.232(4) (2x) |
|  |  | O18 – Mo7 | 2.224(8) |

|  | angle [°] |  | angle [°] |
|---|---|---|---|
| μ$_2$-O |  |  |  |
| Mo1 – O4 – Mo2 (2x) | 107.5(2) | Mo4 – O15 – Mo6 (2x) | 105.1(2) |
| Mo1 – O5 – Mo3 (2x) | 110.5(3) | Mo4 – O16 – Mo4 | 107.6(3) |
| Mo2 – O9 – Mo3 (2x) | 108.2(2) | Mo5 – O19 – Mo5 | 112.6(3) |
|  |  | Mo5 – O20 – Mo7 (2x) | 112.6(2) |
| μ$_3$-O |  |  |  |
| Mo1 – O2 – Mo2 (2x) | 104.5(2) | Mo4 – O14 – Mo4 | 104.6(3) |
| Mo2 – O2 – Mo3 (2x) | 103.5(2) | Mo4 – O14 – Mo6 (2x) | 104.7(2) |
| Mo3 – O2 – Mo1 (2x) | 103.2(2) | Mo5 – O18 – Mo5 | 102.6(3) |
|  |  | Mo5 – O18 – Mo7 (2x) | 101.1(2) |

**Table II.** Upper part: Relations between the exchange and Zeeman energies (g=2.25) to explain the magnetization steps observed in the compound {$Ni_4Mo_{12}$}. Lower part: Exchange parameters (J, J') and standard deviation $\chi^2$ resulting from a least-squares fit to the high-field magnetization data [3].

| Model | Exchange energy | | | | | | | | Zeeman energy |
|---|---|---|---|---|---|---|---|---|---|
| | 1 (x>1) | 2a (x<1) | 2a (x>1) | 2b (x>1) | 3a (x<1) | 3a (x>1) | 3b (x<1) | 3b (x>1) | $g\mu_B H/k_B$ [K] |
| Step 1 | -2J | -2J' | -2J | -2J' | 2J-4J' | -4J+2J' | -J-J' | -J-J' | 6.8(6) |
| Step 2 | -4J | -2J-2J' | -4J | -2J-2J' | 2J-6J' | -2J-2J' | -2.7J-1.3J' | -1.3J-2.7J' | 13.4(1.2) |
| Step 3 | -4J-2J' | -6J | -2J-4J' | -2J-4J' | -6J | -2J-4J' | -4.8J-1.2J' | -1.2J-4.8J' | 30.4(1.2) |
| Step 4 | -4J-4J' | -8J | -2J-6J' | -4J-4J' | -8J | -2J-6J' | -6.8J-1.2J' | -1.2J-6.8J' | 48.4(3.0) |
| J [K] | -3.36(5) | -5.16(44) | -3.35(26) | -4.5(1.9) | -5.26(33) | -4.21(57) | -6.45(82) | 0.5(1.9) | |
| J' [K] | -8.59(11) | -3.03(69) | -6.20(34) | -4.5(1.9) | -4.21(27) | -5.19(75) | 0.5(1.9) | -6.45(82) | |
| $\chi^2$ | 0.31 | 3.53 | 1.77 | 7.30 | 2.69 | 5.50 | 3.58 | 3.58 | |

**FIGURE CAPTIONS**

**Figure 1:**

View of the $[Mo_{12}O_{28}(\mu_2\text{-}OH)_9(\mu_3\text{-}OH)_3\{Ni(H_2O)_3\}_4]$ molecule. $\mu_2$–oxo and $\mu_3$–oxo ions are emphasized in yellow color. H atoms are omitted.

**Figure 2:**

(a) Schematic sketch of the $[Ni_4Mo_{12}]$ molecule. All edges of the tetrahedron are bridged by dimolybdate groups, only one is shown for clarity. (b) Detailed view of the bridging geometry $(H_2O)_3Ni[Mo_2O_{10}]Ni(H_2O)_3$.

**Figure 3:**

High-field differential magnetization $\Delta M/\Delta H$ of the compound $\{Ni_4Mo_{12}\}$. The circles denote the experimental data taken from Ref. 3. The full and dashed lines correspond to the calculations based on the Hamiltonians (5) and (14) with the model parameters of Eqs (13) and (15), respectively.

**Figure 4:**

Energy spectra of neutrons scattered from the polycrystalline tetrameric $Ni^{2+}$ compound $\{Ni_4Mo_{12}\}$ as a function of both temperature T (left panel) and modulus of the scattering vector $\mathbf{Q}$ (right panel), using an incident neutron energy of $E_0$=5.93 meV.

**Figure 5:**

Energy spectra of neutrons scattered from polycrystalline {Ni$_4$Mo$_{12}$} with an incident neutron energy $E_0$=3.27 meV. (a) Measurements performed in the present work. (b) Background corrected data taken from Ref. 5.

**Figure 6:**

Double-logarithmic plot of the temperature dependence of the transitions centered at 0.5, 1.1, 1.35 and 1.7 meV. The full and dashed lines denote the Boltzmann population factors $n_0$ and $n_1$, respectively, which are calculated from the energy level sequence derived for {Ni$_4$Mo$_{12}$} (see Section IV.C).

**Figure 7:**

Q-dependence of the ground-state transitions observed at 0.5 and 1.7 meV (upper panel) as well as at 1.35 meV (lower panel).

**Figure 8:**

Energy levels of Ni$^{2+}$ tetramers normalized to J (assuming antiferromagnetic coupling J<0) as a function of the ratio x=J'/J for models 1 and 2a. The lowest states of a given S value (S=0,1,2,3,4) are marked by bold lines. The inserts show the exchange couplings (J, J') associated within the tetrahedra of Ni$^{2+}$ ions.

**Figure 9:**

Energy levels of Ni$^{2+}$ tetramers normalized to J (assuming antiferromagnetic coupling J<0) as a function of the ratio x=J'/J for models 2b, 3a, and 3b. The bold lines and the inserts are as in Fig. 8.

**Figure 10:**

Difference energy spectrum ($T_1$=1.5 K, $T_2$=10 K) observed for polycrystalline {Ni$_4$Mo$_{12}$} with incoming neutron energy $E_0$=5.93 meV. The line corresponds to the calculated spectrum with the model parameters of Eqs (13) and (15) (the two calculated spectra do not differ from each other within the thickness of the line). The open symbols denote the data excluded from the fit.

**Figure 11:**

Difference energy spectra ($T_1$, $T_2$) observed for polycrystalline {Ni$_4$Mo$_{12}$} with incoming neutron energy $E_0$=3.27 meV. (a) Measurements performed in the present work with $T_1$=1.5 K and $T_2$=10 K. The line and the symbols are as in Fig. 10. (b) Measurements reported in Ref. 5 with $T_1$=2.4 K and $T_2$=9.3 K. The line corresponds to the calculated spectrum with the model parameters of Eq. (16).

**Figure 12:**

Energy level sequence of the spin states associated with the Ni$^{2+}$ tetramers in {Ni$_4$Mo$_{12}$} corresponding to the model parameters of Eq. (13). The lowest states of a given S value are marked with bold lines. The full, dashed and dotted arrows mark the relevant low-energy transitions with squared transition matrix elements $P_{i \to j}$ >2.5, 2.5 > $P_{i \to j}$ >1 and $P_{i \to j}$ <1, respectively.

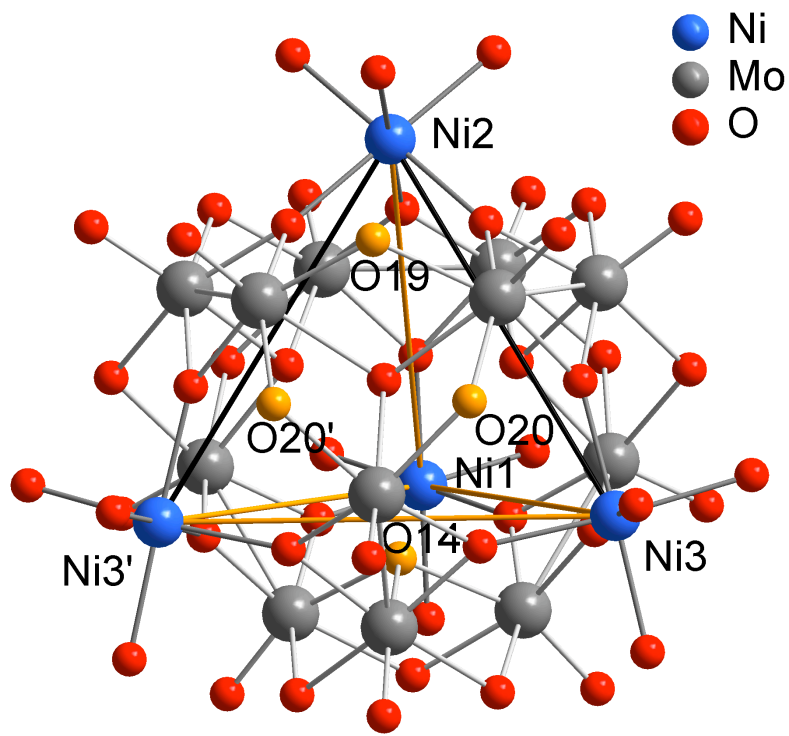

**Figure 1:**

View of the [Mo$_{12}$O$_{28}$(µ$_2$-OH)$_9$(µ$_3$-OH)$_3${Ni(H$_2$O)$_3$}$_4$] molecule. µ$_2$–oxo and µ$_3$–oxo ions are emphasized in yellow color. H atoms are omitted.

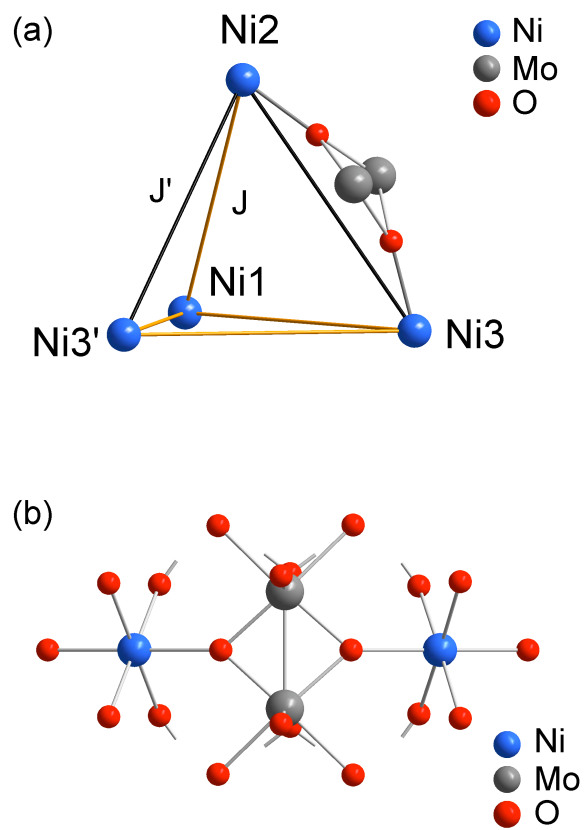

**Figure 2:**

(a) Schematic sketch of the [Ni$_4$Mo$_{12}$] molecule. All edges of the tetrahedron are bridged by dimolybdate groups, only one is shown for clarity. (b) Detailed view of the bridging geometry (H$_2$O)$_3$Ni[Mo$_2$O$_{10}$]Ni(H$_2$O)$_3$.

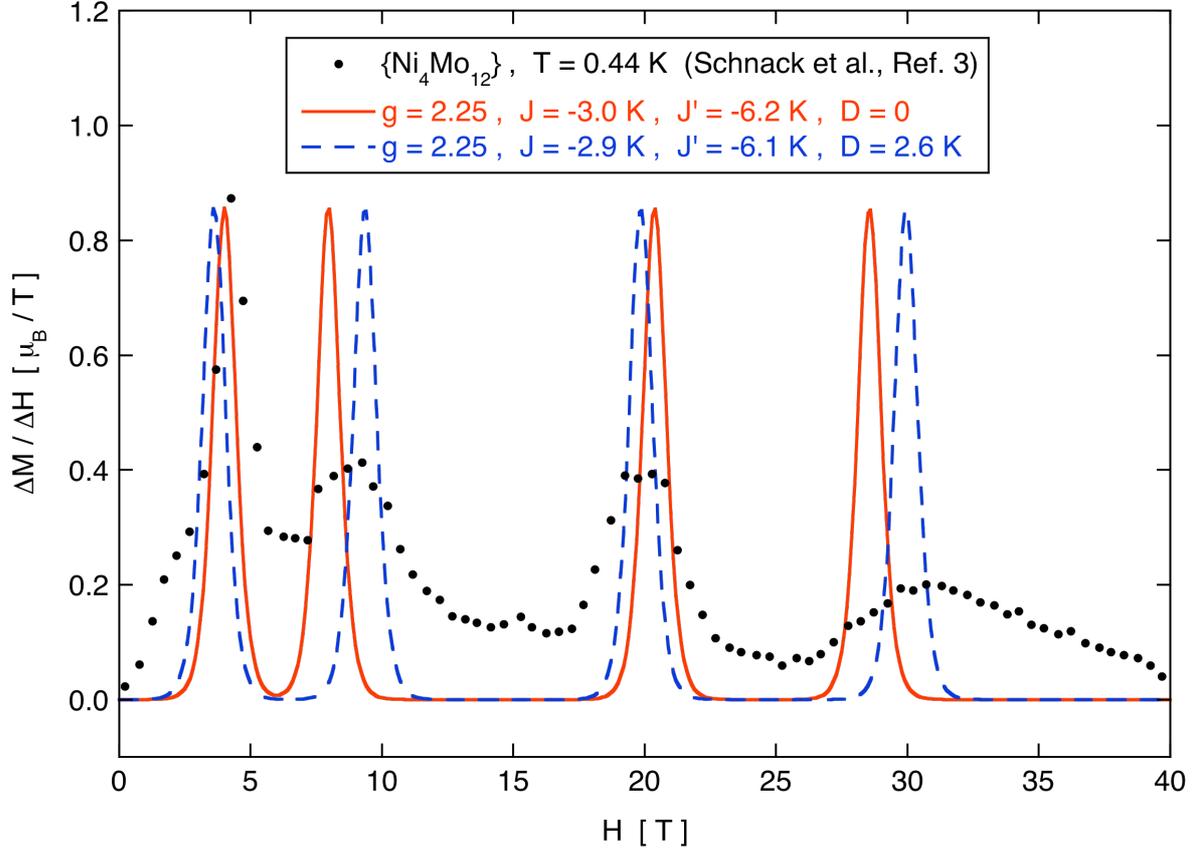

**Figure 3:**

High-field differential magnetization ΔM/ΔH of the compound {Ni$_4$Mo$_{12}$}. The circles denote the experimental data taken from Ref. 3. The full and dashed lines correspond to the calculations based on the Hamiltonians (5) and (14) with the model parameters of Eqs (13) and (15), respectively.

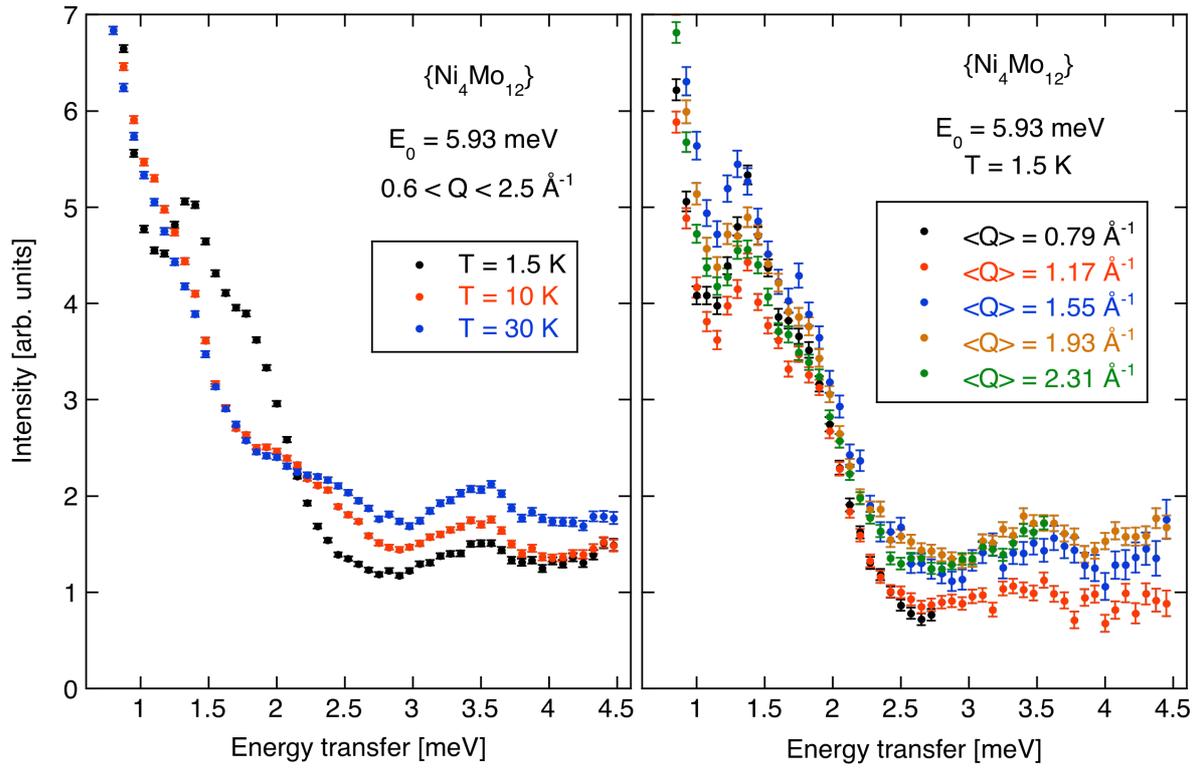

**Figure 4:**

Energy spectra of neutrons scattered from the polycrystalline tetrameric Ni$^{2+}$ compound {Ni$_4$Mo$_{12}$} as a function of both temperature T (left panel) and modulus of the scattering vector **Q** (right panel), using an incident neutron energy of E$_0$=5.93 meV.

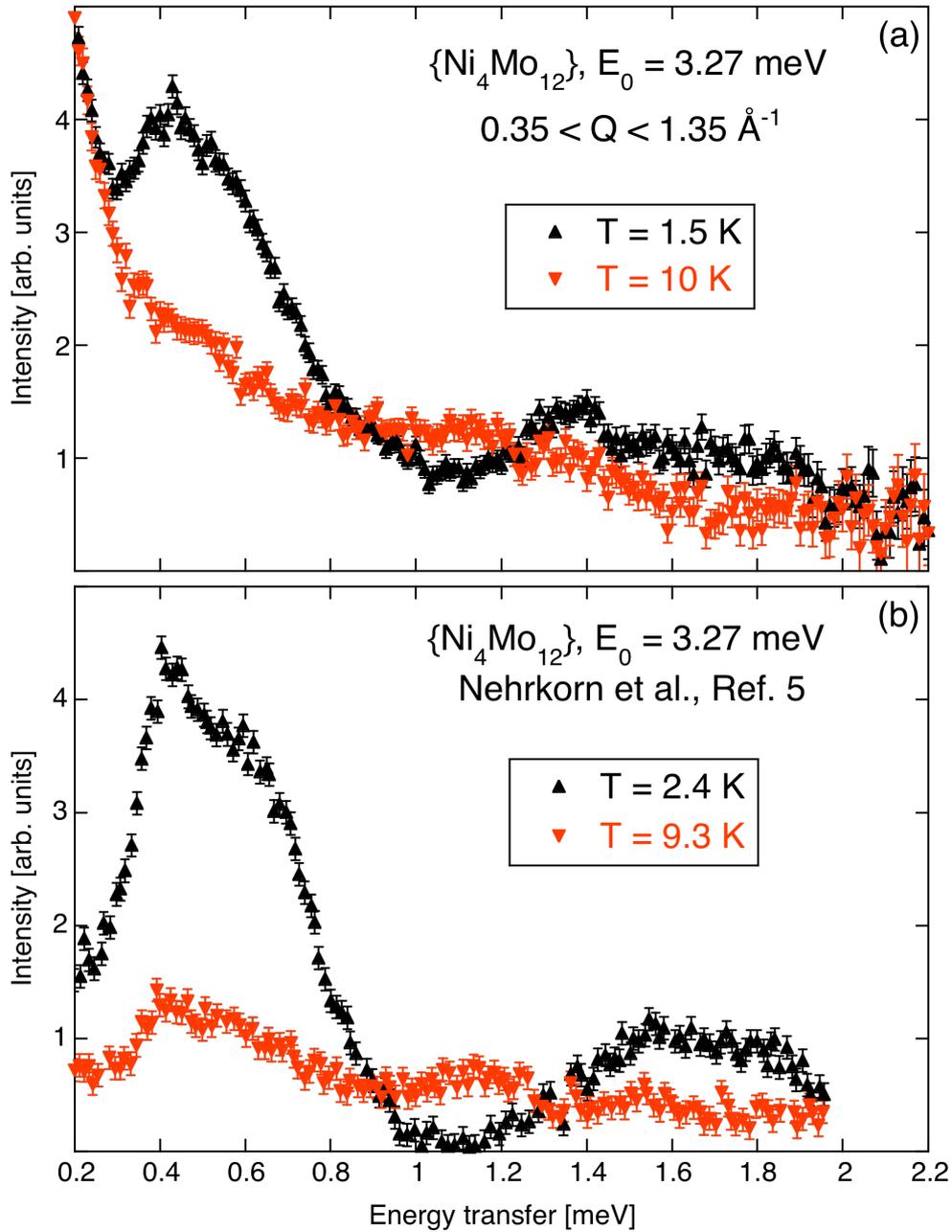

**Figure 5:**

Energy spectra of neutrons scattered from polycrystalline {Ni$_4$Mo$_{12}$} with an incident neutron energy $E_0$=3.27 meV. (a) Measurements performed in the present work. (b) Background corrected data taken from Ref. 5.

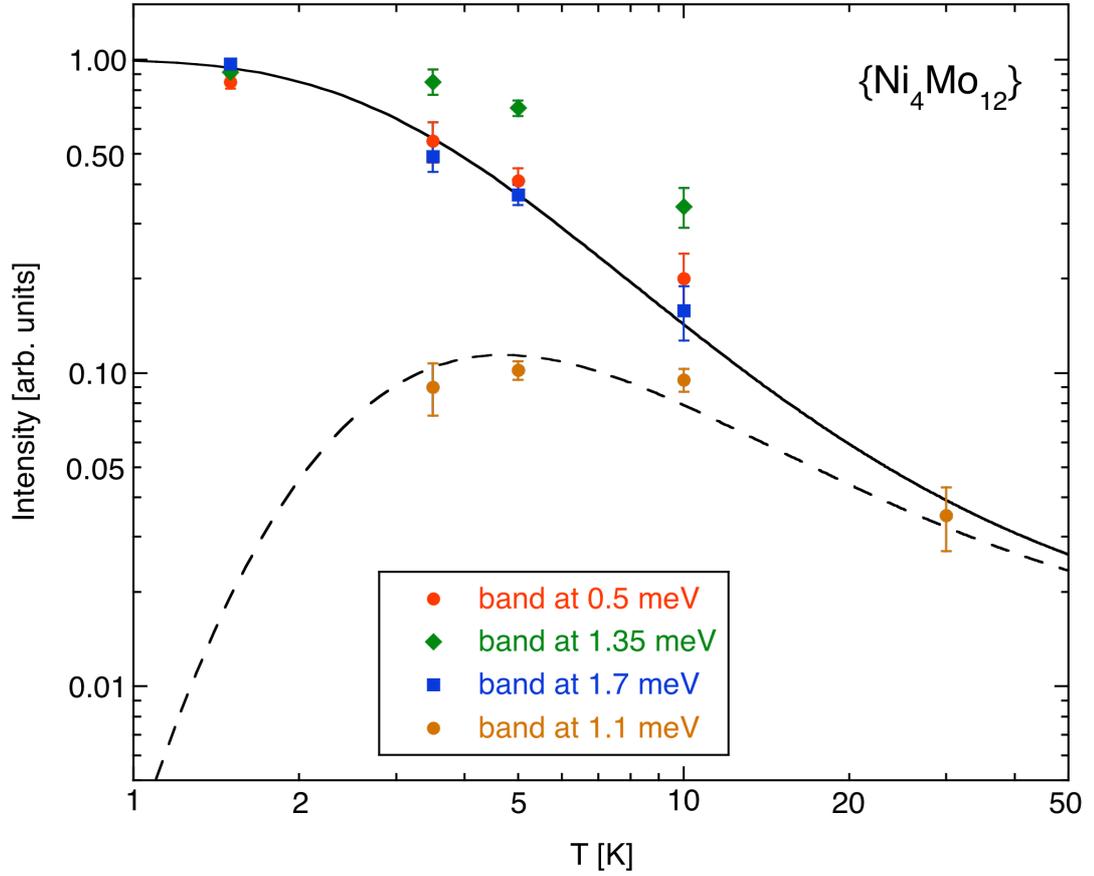

**Figure 6:**

Double-logarithmic plot of the temperature dependence of the transitions centered at 0.5, 1.1, 1.35 and 1.7 meV. The full and dashed lines denote the Boltzmann population factors $n_0$ and $n_1$, respectively, which are calculated from the energy level sequence derived for {Ni$_4$Mo$_{12}$} (see Section IV.C).

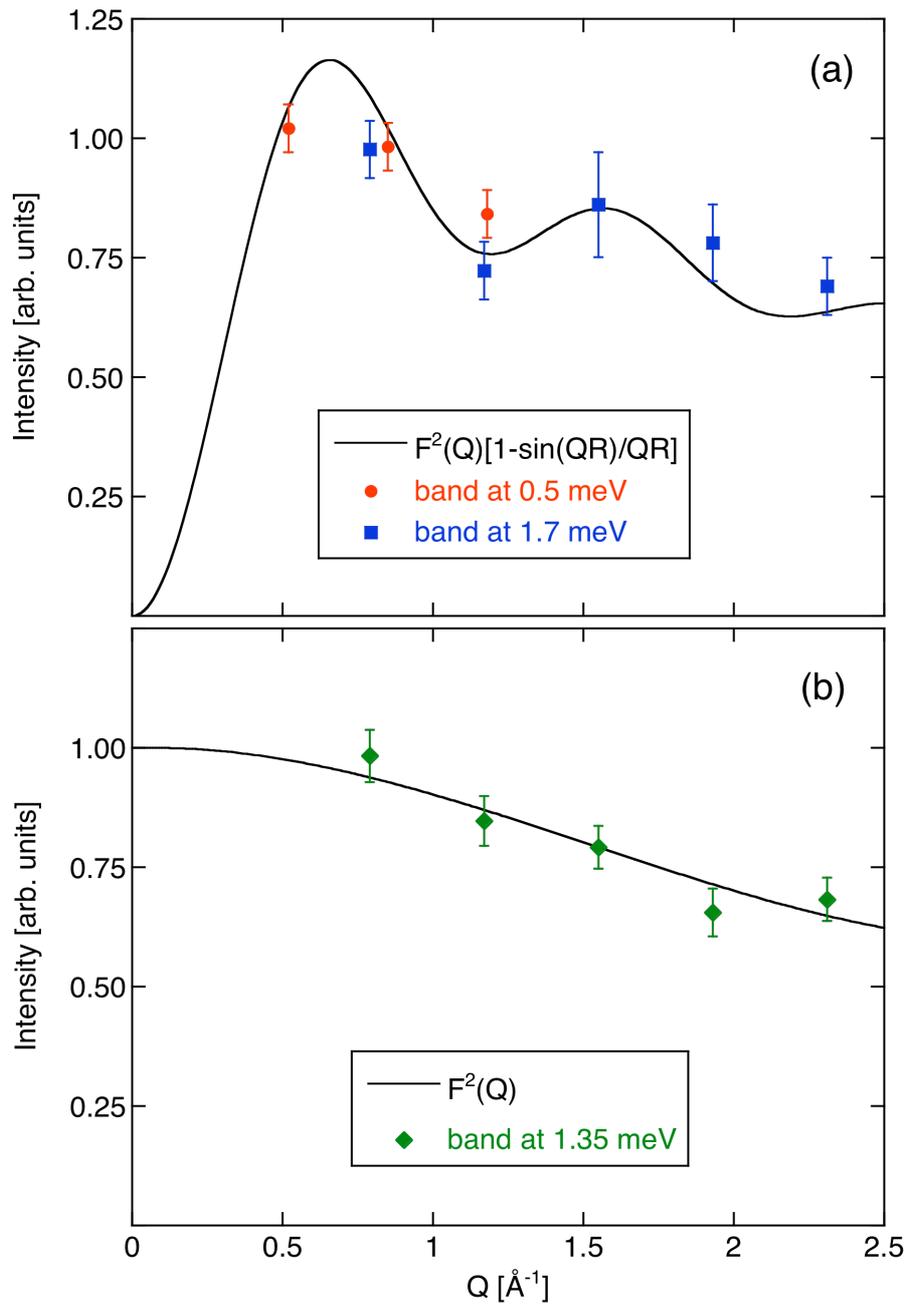

**Figure 7:**

Q-dependence of the ground-state transitions observed at 0.5 and 1.7 meV (upper panel) as well as at 1.35 meV (lower panel).

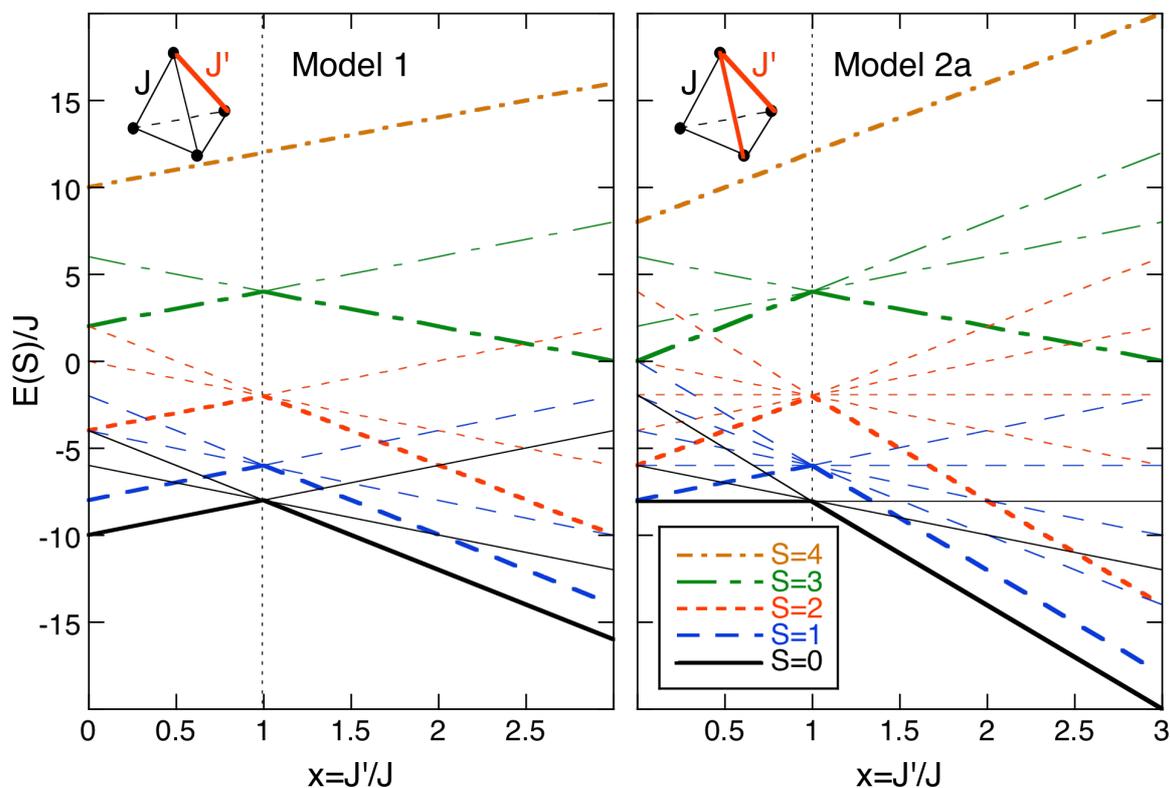

**Figure 8:**

Energy levels of $Ni^{2+}$ tetramers normalized to J (assuming antiferromagnetic coupling J<0) as a function of the ratio x=J'/J for models 1 and 2a. The lowest states of a given S value (S=0,1,2,3,4) are marked by bold lines. The inserts show the exchange couplings (J, J') associated within the tetrahedra of $Ni^{2+}$ ions.

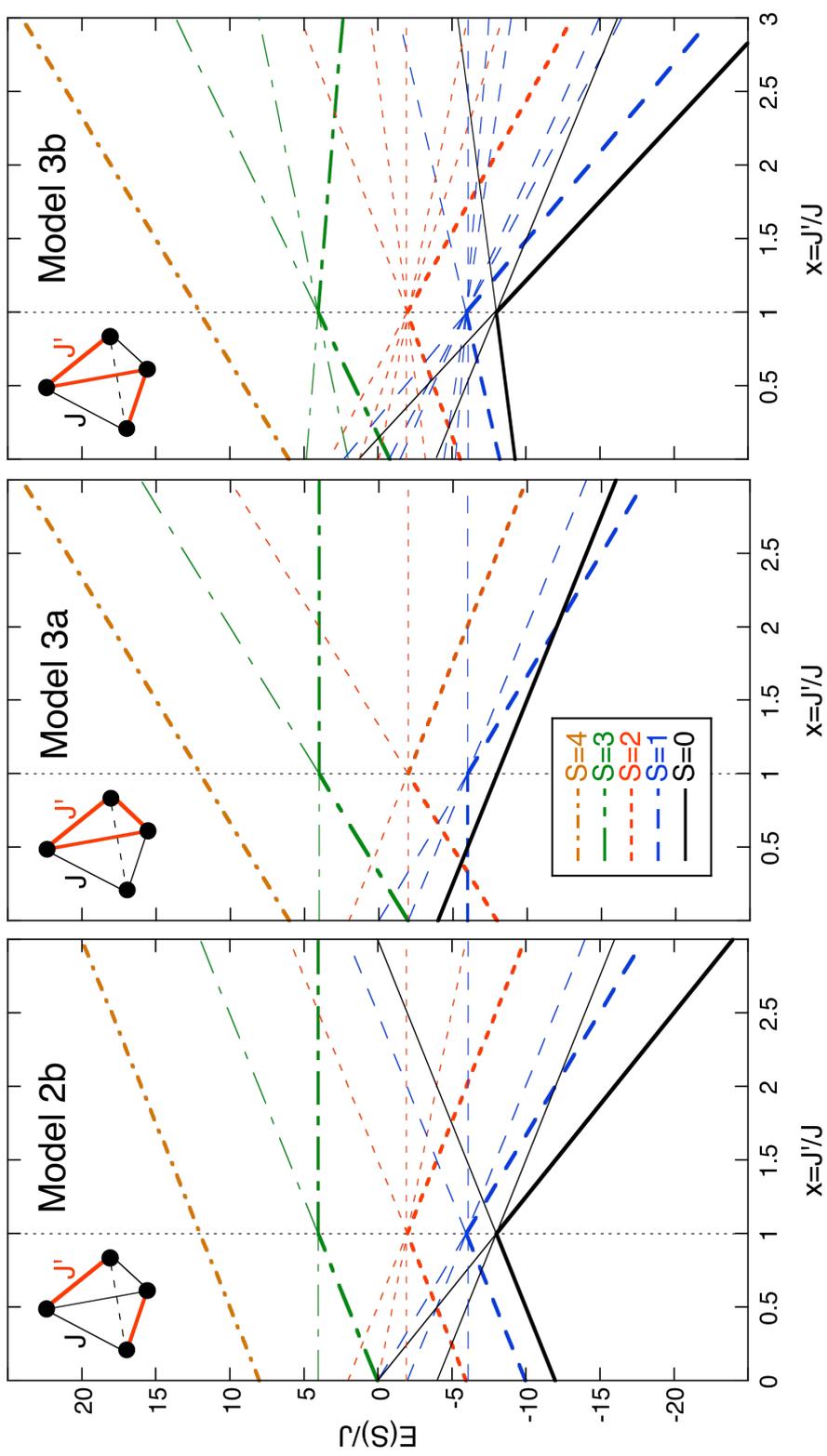

**Figure 9:**

Energy levels of Ni$^{2+}$ tetramers normalized to $J$ (assuming antiferromagnetic coupling $J<0$) as a function of the ratio $x=J'/J$ for models 2b, 3a, and 3b. The bold lines and the inserts are as in Fig. 8.

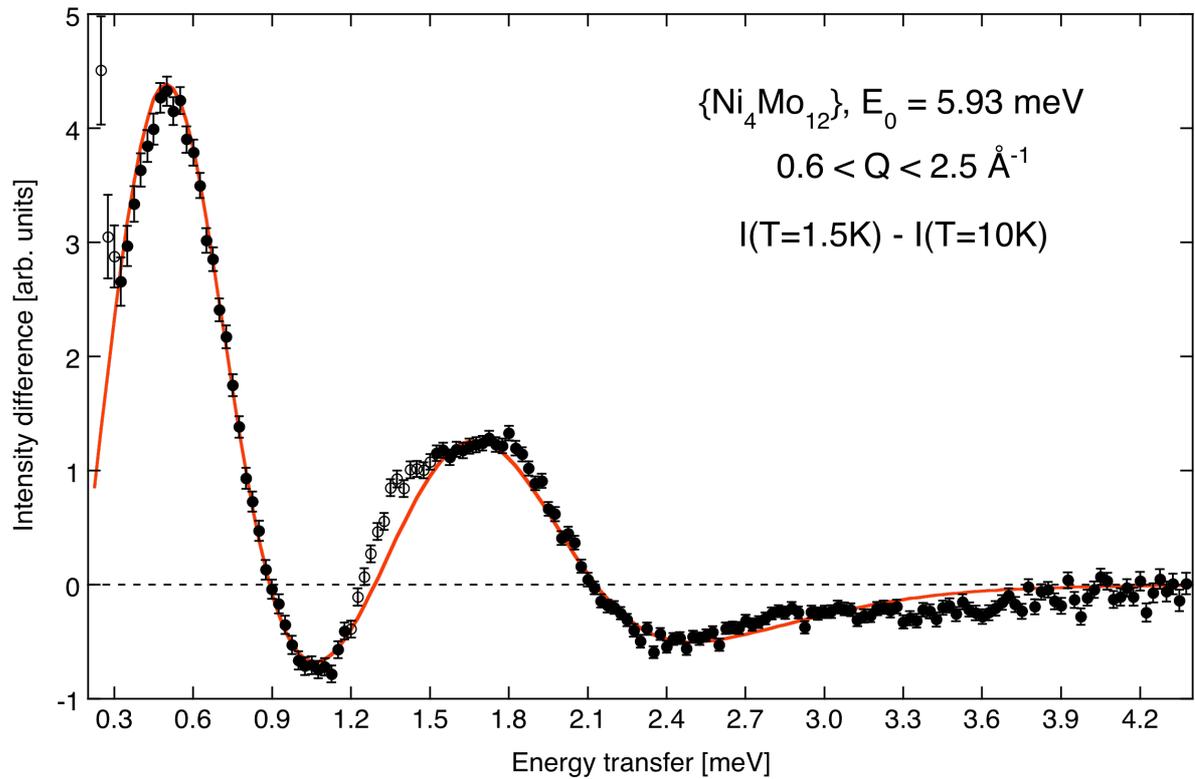

**Figure 10:**

Difference energy spectrum ($T_1$=1.5 K, $T_2$=10 K) observed for polycrystalline {Ni$_4$Mo$_{12}$} with incoming neutron energy $E_0$=5.93 meV. The line corresponds to the calculated spectrum with the model parameters of Eqs (13) and (15) (the two calculated spectra do not differ from each other within the thickness of the line). The open symbols denote the data excluded from the fit.

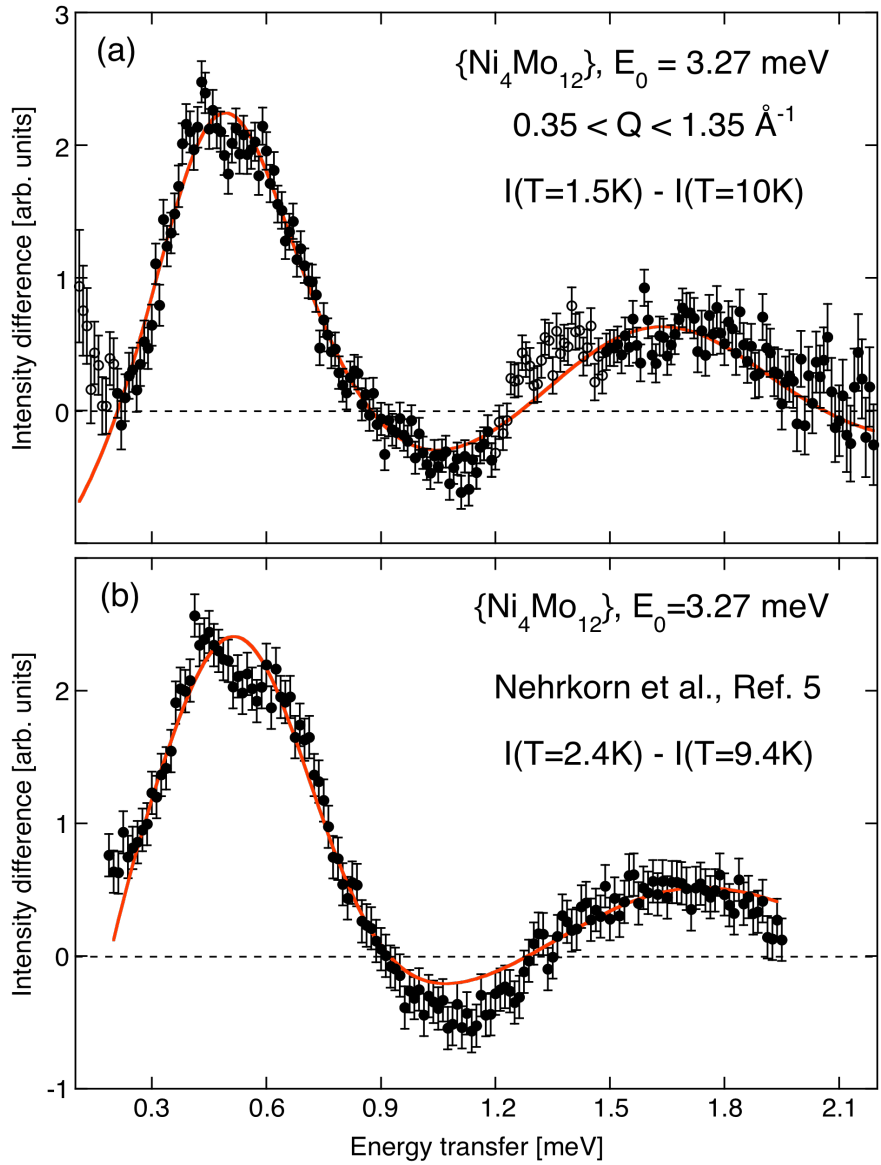

**Figure 11:**

Difference energy spectra ($T_1$, $T_2$) observed for polycrystalline {Ni$_4$Mo$_{12}$} with incoming neutron energy $E_0$=3.27 meV. (a) Measurements performed in the present work with $T_1$=1.5 K and $T_2$=10 K. The line and the symbols are as in Fig. 10. (b) Measurements reported in Ref. 5 with $T_1$=2.4 K and $T_2$=9.3 K. The line corresponds to the calculated spectrum with the model parameters of Eq. (16).

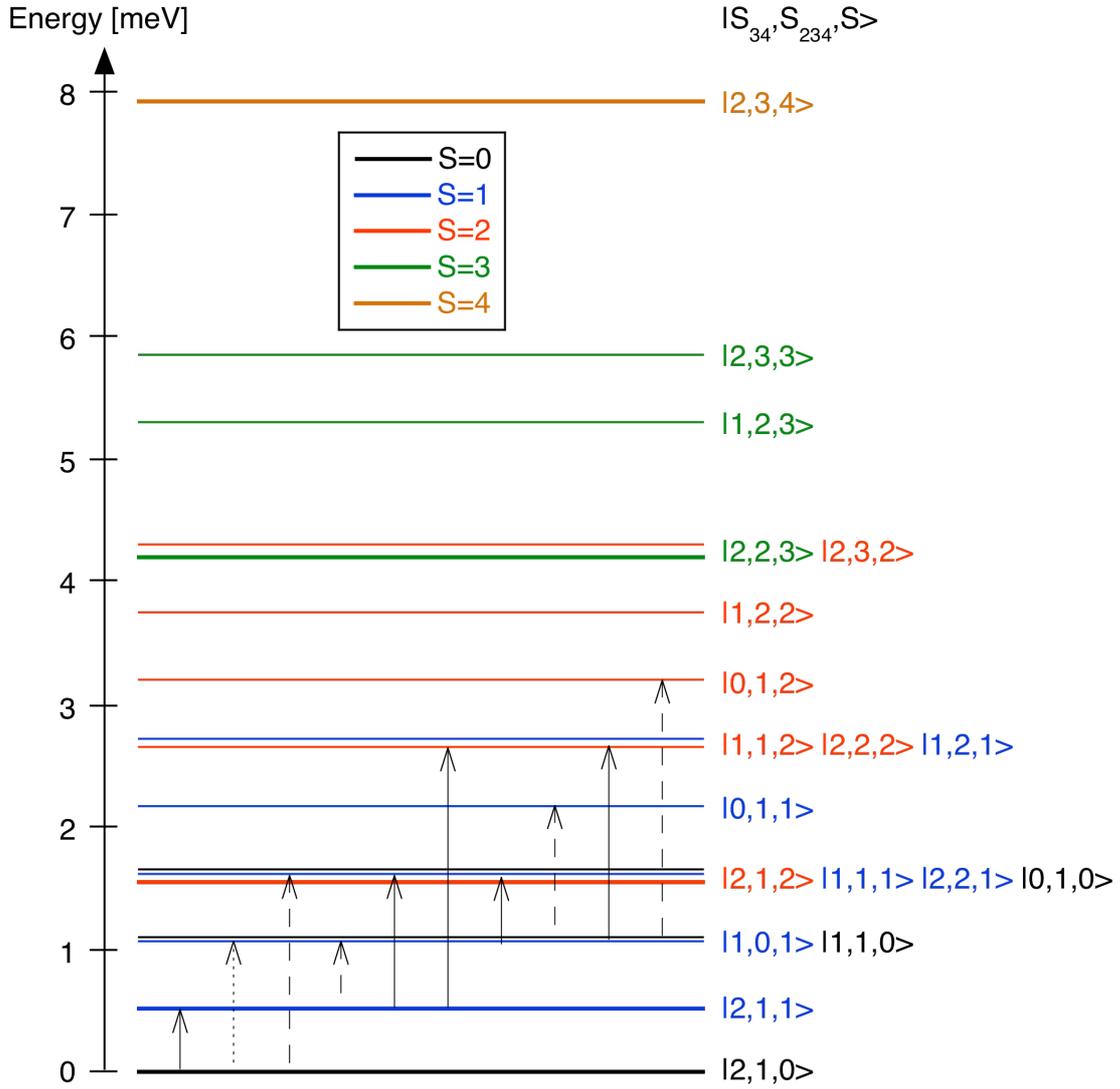

**Figure 12:**

Energy level sequence of the spin states associated with the Ni$^{2+}$ tetramers in {Ni$_4$Mo$_{12}$} corresponding to the model parameters of Eq. (13). The lowest states of a given S value are marked with bold lines. The full, dashed, and dotted arrows mark the relevant low-energy transitions with squared transition matrix elements P$_{i \rightarrow j}$ >2.5, 2.5 >P$_{i \rightarrow j}$ >1, and P$_{i \rightarrow j}$ <1, respectively.